\documentstyle[aps,prd]{revtex}
\input{epsf}

\begin{document}

\draft

\title{Cauchy-perturbative matching and outer
       boundary conditions I: Methods and tests}

\author{M.~E.~Rupright,$^{1}$ A.~M.~Abrahams,$^{2,3,4}$ and 
	L.~Rezzolla$^{2,3}$}

\address{$^{1}$Department of Physics and Astronomy, University of
         North Carolina, Chapel Hill, North Carolina 27599-3255}
\address{$^{2}$Department of Physics, University of Illinois at
	Urbana-Champaign, Urbana, Illinois 61801}
\address{$^{3}$NCSA, University of Illinois at
	Urbana-Champaign, Urbana, Illinois 61801}
\address{$^{4}$J.~P. Morgan, 60 Wall St., New York, New York 10260}

\date{\today}
\maketitle

\begin{abstract}
	We present a new method of extracting gravitational radiation
from three-dimensional numerical relativity codes and providing outer
boundary conditions. Our approach matches the solution of a Cauchy
evolution of Einstein's equations to a set of one-dimensional linear
wave equations on a curved background. We illustrate the mathematical
properties of our approach and discuss a numerical module we have
constructed for this purpose. This module implements the perturbative
matching approach in connection with a generic three-dimensional
numerical relativity simulation. Tests of its accuracy and
second-order convergence are presented with analytic linear wave data.
\end{abstract}


\section{Introduction}
\label{sec:intro}

	An important goal of numerical relativity is to compute the
gravitational waveforms generated by systems of compact astrophysical
objects such as binary black holes or binary neutron stars.  With the
prospect that gravitational wave detectors such as LIGO, VIRGO and GEO
will be on-line in the next few years, it is crucial to study
numerical relativistic simulations of events which might be observable
by these detectors. Such calculations are important not only because
they could provide signal templates which would considerably increase
the probability of detection, but also because the comparison of such
templates with the observations may provide essential astrophysical
information on the nature of the emitting sources. The purpose of the
Binary Black Hole ``Grand Challenge'' {\it Alliance} \cite{BBHGCA}, a
multi-institutional collaboration in the United States, is to study
the inspiral coalescence of the most significant source of signals for
the interferometric gravity wave detectors: a binary black hole
system.

	Central to the goal of determining waveforms generated by
astrophysical systems is the need for accurate techniques which
compute asymptotic waveforms from numerical relativity simulations on
three-dimensional (3D) spacelike hypersurfaces with finite extents. In
general, the computational domain cannot be extended to the distant
wave zone \cite{Thorne}, where the geometric optics approximation
is valid. Indeed, computational resource limitations require that the
outer boundary of such simulations lies rather close to the highly
dynamical and strong field region, where backscatter of waves off
curvature can be significant. As a result, it is imperative to develop
techniques which can ``extract'' the gravitational waves generated
by the simulation and evolve them out to the distant wave zone where
they assume their asymptotic form.

	While the problem of radiation extraction is important for
computing observable waveforms from numerical simulations, careful
implementation of outer boundary conditions is also crucial for 
maintaining the integrity of the simulations themselves, as poorly 
implemented boundary conditions are a likely source of numerical
instabilities.  These outer boundary conditions are also decisive in
framing the desired physical context for the simulation, {\it e.g.}, 
an isolated source in an asymptotically flat spacetime.  For typical
applications, we can summarize the requirements of a
radiation-extraction/outer-boundary module as: {\it (a)} supporting
stable evolution of Einstein's equations, {\it (b)} minimizing
spurious (numerical) reflection of radiation at the boundary, {\it
(c)} providing accurate and numerically convergent approximations to
the gravitational waveforms that would be observed in the wave zone
surrounding an isolated source, {\it (d)} incorporating effects of
radiation reflection off background curvature outside the numerical
boundary when appropriate (for example when the outer boundary is in a
strong field region).

	In this paper we present a new method for extracting
gravitational waveforms from a 3D numerical relativity code while {\it
simultaneously} imposing outer boundary conditions. Our approach is
motivated by earlier investigations of gauge-invariant extraction
techniques \cite{AbrahamsEvans}, but promises to be more generally
applicable in cases where the background curvature is significant near
the outer boundary of the computational 3D grid. Our method matches a
full 3D Cauchy solution of Einstein's equations on spacelike
hypersurfaces with a perturbative one-dimensional (1D) solution in a
region where the waveforms can be treated as linear perturbations on a
spherically symmetric curved background \footnote{An alternative 
		approach to the problem of wave 
		extraction and outer boundary conditions has been
		developed to match the Cauchy 
		solution to solutions
		on characteristic hypersurfaces \cite{Characteristic}.}.

	The plan of this paper is as follows: in Section
\ref{sec:analysis}, we describe the mathematical basis of our method
and derive the linearized radial wave equations which account for the
evolution of the gravitational waves in the perturbative region of the
spacetime. In Section \ref{sec:procimplement}, we discuss the
strategies for the numerical solution of the above equations and
present a numerical code we have constructed which represents a
general module implementing our extraction/outer boundary method in
conjunction with a 3D numerical relativity simulation. In this paper
we focus on tests of the outer boundary module as a self-contained
unit using analytic solutions.  In companion papers we focus on tests
of the module in the practical context of a typical application.  In
\cite{AlliancePRL}, we have presented tests of this outer boundary
module in conjunction with the 3D {\it ``interior code''} of the
Alliance which evolves Einstein equations in the standard
``$3+1$'' form (as presented in \cite{York79}). A
more thorough discussion of these results is forthcoming \cite{prd2}.

\section{The Cauchy-perturbative matching method}
\label{sec:analysis}

	Einstein's equations are highly nonlinear and when 
spacetime is characterized by rapidly varying strong fields,
the full 3D nonlinear equations must be used. Outside of 
an isolated region of this kind, however, a perturbative approximation, 
in which gravitational data are treated as linear perturbations of an 
exact solution to Einstein's equations, may be valid. In this 
perturbative region a linearized approximation to Einstein's equations 
could then exploited to simplify the evolution of gravitational data.

\begin{figure}[t]
\epsfxsize=10cm
\begin{center}
\leavevmode
\epsffile{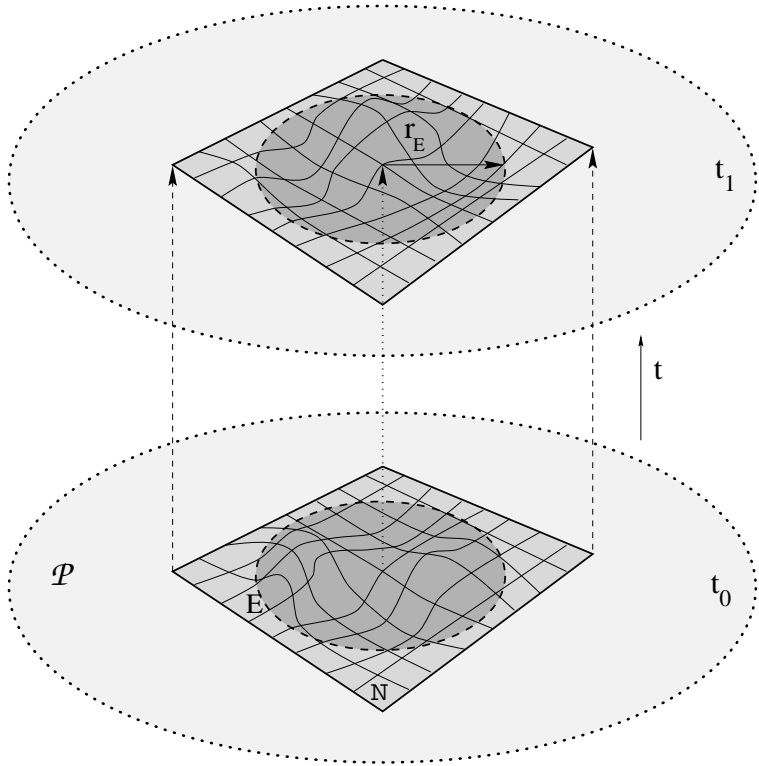}
\vskip 1.0truecm 
\end{center}
\caption[fig1]{\label{fig:schematic}
	Schematic of matching procedure for two successive timeslices 
	(one dimension is suppressed). {\tt N} is the 3D
	computational domain. The dark shaded region shows the 
	strong field highly dynamical region in {\tt N}.
	The medium and light shaded annular region represents the
	``perturbative region'', ${\cal P}$, where the linearized
	equations are evolved on a 1D grid {\tt L} (not shown).}
\end{figure}

	The idea behind a Cauchy-perturbative matching approach is to
supplement the computationally expensive evolution of the full
Einstein equations with the comparatively simpler evolution of the
linearized equations in a perturbative region. Figure
\ref{fig:schematic} provides a schematic picture of the
Cauchy-perturbative matching approach. The square region covered by
the grid represents the 3D computational domain {\tt N } (one
dimension is suppressed) on which Cauchy evolution of the full
Einstein equations is computed.  The dark central area in {\tt N}
includes the strong field highly dynamical region, where the nonlinear
Einstein equations must be solved.  The medium and light shaded
annular area, ${\cal P}$, represents the perturbative region. Anywhere
in the (medium shaded) intersection of {\tt N} and ${\cal P}$, we can
place an {\it extraction} 2-sphere {\bf E}, of radius $r_{_E}$, where
the gravitational field information is read out. This information is
then evolved (by means of the linearized Einstein equations) in ${\cal
P}$, which ranges from {\bf E} out to a large distance (shown as a
dotted circle) where the asymptotic waveforms can be identified. Outer
boundary data for {\tt N} can be constructed from perturbative data in
the intersection of {\tt N} and ${\cal P}$.

	Previous investigations \cite{AbrahamsEvans} achieved the
desired perturbative simplification by matching the nonlinear solution
onto analytic solutions of Einstein's equations linearized on a
Minkowski background. Further simplification was achieved by
decomposing perturbative data in a multipole expansion. We extend this
approach to cases where curvature is significant by choosing as our
approximation a linearization of Einstein's equations on a
Schwarzschild background. In principal one could generalize further
to a Kerr background, and this will be the subject of future work.
We also decompose perturbative data on this background with a multipole 
expansion, reducing the 3D linearized equations to a set of 1D 
equations for each multipole mode. This reduction allows us to evolve 
data everywhere in ${\cal P}$ on a one-dimensional grid, {\tt L}.
It is important to note that all of the 3-dimensional tensor data
in ${\cal P}$ can be reconstructed from the multipole amplitudes
on {\tt L}. Our method, therefore, is to match a computationally 
expensive evolution of the full Einstein equations onto a considerably 
less expensive evolution on a 1D grid in a region where background 
curvature is still significant.

\subsection{Hyperbolic formulation}
\label{subsec:hyper}

	Rather than characterize radiation asymptotically in terms
of certain variables constructed from the metric \cite{AbrahamsEvans},
we use a new approach which characterizes radiation in terms of the
extrinsic curvature. This is made possible by a recently
developed spatially gauge-covariant hyperbolic formulation of general
relativity. This system is constructed from first derivatives of the
spacetime Ricci tensor \cite{CBR,CBY,AACBY} and may therefore 
appropriately be called the ``Einstein--Ricci'' system.

The Einstein--Ricci equations are obtained from the ``$3+1$'' form
of the metric,
\begin{equation}
ds^2 = -N^2 dt^2 + g_{i j} (dx^i + \beta^i dt) (dx^j + \beta^j dt) \ ,
\end{equation}
where $N$ is the lapse function, $\beta^i$ is the shift vector, and
$g_{i j}$ is the spatial metric in the slice $\Sigma$.  An appropriate
time derivative operator that evolves spatial quantities along the
normal to the slice $\Sigma$ is
\begin{equation}
\hat\partial_0 = \partial_t - {\cal L}_\beta \ ,
\end{equation}
where ${\cal L}_\beta$ is the Lie derivative along the 
shift vector in $\Sigma$. 

The extrinsic curvature $K_{i j}$ of $\Sigma$ can be defined by
\begin{equation}
\hat\partial_0 g_{i j} = -2 N K_{i j} \ ,
\label{eq:defK}
\end{equation}
which serves also as the evolution equation for the spatial metric.
By working out the expression $\Omega_{i j} \equiv \hat\partial_0 R_{i
j} - 2 \overline{\nabla}_{( i} R_{j ) 0}$ in $3+1$ form, where $R_{i
j}$ and $R_{j 0}$ are components of the spacetime Ricci tensor and
$\overline{\nabla}_i$ denotes the spatial covariant derivative, we
find a wave-like equation which governs the evolution of $K_{i j}$:

\begin{equation}
-N \hat{\mbox{\kern-.0em\lower.3ex\hbox{$\Box$}}} K_{i j} = 
	J_{i j} + S_{i j} - \Omega_{i j} \ ,
\label{eq:boxK}
\end{equation}
where the physical wave operator for arbitrary shift is
$\hat{\mbox{\kern-.0em\lower.3ex\hbox{$\Box$}}} \equiv -N^{-1}
\hat\partial_0 N^{-1} \hat\partial_0 + \overline{\nabla}_k
\overline{\nabla}^k$. Equation (\ref{eq:boxK}) is an identity until we
substitute the Einstein equations $R_{\alpha \beta} = 
8 \pi ( T_{\alpha \beta} - \frac{1}{2} T^{\lambda}_{\lambda}
g_{\alpha \beta} )$ into $\Omega_{i j}$ ( $G = c = 1$ ).

	The detailed form of the right hand side of (\ref{eq:boxK}) can be
found in \cite{CBY,AACBY}; the present conventions are those in 
\cite{AACBY}. Here we simply point out that $\Omega_{i j}$ has become
a matter source that is zero here, $J_{i j}$ is the nonlinear 
self-interaction term in $3+1$ form, 
and $S_{i j}$ is a slicing-dependent term that must involve fewer than 
second derivatives of $K_{i j}$ to render (\ref{eq:boxK}) a true
(hyperbolic) wave equation. A simple way to satisfy the restriction on
$S_{i j}$ is to invoke the harmonic slicing condition
\begin{equation}
\hat\partial_0 N + N^2 H = 0 , 
\label{eq:harmonic}
\end{equation}
where $H$ is the trace of $K_{i j}$, and from which follows 
$S_{i j}=0$.

	For appropriate choice of initial data \cite{CBY,AACBY},
equations (\ref{eq:defK}), (\ref{eq:boxK}), and (\ref{eq:harmonic})
represent the dynamical part of Einstein's equations.  Combining
them we obtain a quasi-diagonal hyperbolic equation for $g_{i j}$, 
with principal (highest-order) part 
$\hat{\mbox{\kern-.0em\lower.3ex\hbox{$\Box$}}} \hat\partial_0$.  
Hence (\ref{eq:defK}), (\ref{eq:boxK}), and
(\ref{eq:harmonic}) may be said to give the ``third-order'' form of
the Einstein--Ricci system.

We note that the third-order Einstein--Ricci system can also be 
cast into a first-order symmetric hyperbolic form 
\cite{CBY,AACBY,AACBY2}.~\footnote{In \cite{AACBY}, 
the equation for $\hat\partial_0
\overline{\Gamma}^i_{j k}$ was inadvertently omitted. See
\cite{CBY,AACBY2}.}  It also possesses a higher order form (the
``fourth-order Einstein--Ricci system''), essentially a wave equation
for $(\hat\partial_0 K_{i j})$, obtained from $\hat\partial_0
\Omega_{i j} + \overline{\nabla}_i \overline{\nabla}_j R_{0 0}$
\cite{AACBY2,AACBY3,CBYBanach}. This system has a well-posed Cauchy 
problem and complete freedom in choosing both $\beta^i$ and $N$: 
it has no analog of a slicing term like $S_{i j}$.  This 
fourth-order form is used to develop fully gauge-invariant 
perturbation theory in \cite{WorkingPaper}.

\subsection{Perturbative Expansion}
\label{subsec:pertexpans}

	The first step in obtaining radial wave equations is to
linearize the hyperbolic Einstein-Ricci equations around a static
Schwarzschild background. We separate the gravitational quantities of
interest into background (denoted by a tilde) and perturbed parts: the
3-metric $g_{i j} = \widetilde{g}_{i j} + h_{i j}$, the extrinsic
curvature $K_{i j} = \widetilde{K}_{i j} + \kappa_{i j}$, the lapse $N
= \widetilde{N} + \alpha$, and the shift vector $\beta^i =
\widetilde{\beta}^i + v^i$.  In Schwarzschild coordinates $(t, r,
\theta, \phi)$, the background quantities are given by

\begin{mathletters}
\label{eq:background}
\begin{eqnarray}
\widetilde{N} &=& \left(1 - \frac{2M}{r}\right)^{1/2} \ ,         \\
\widetilde{g}_{i j} dx^i dx^j &=& \widetilde{N}^{-2} dr^2 +  
		r^2 (d\theta^2 + \sin^2\theta d\phi^2) \ ,         \\
\widetilde{\beta}^i &=& 0 \ ,                           	   \\
\widetilde{K}_{i j} &=& 0 \ ,                             
\end{eqnarray}
\end{mathletters}
while the perturbed quantities have arbitrary angular dependence.
The background quantities satisfy the dynamical equations
$\partial_t \widetilde{g}_{i j} = 0$, $\partial_t \widetilde{N} = 0$,
and thus remain constant for all time. The perturbed quantities,
on the other hand, obey the following evolution equations

\begin{mathletters}
\label{eq:evolution}
\begin{eqnarray}
\partial_t h_{i j} &=& -2 \widetilde{N} \kappa_{i j} + 
		2 \widetilde{\nabla}_{(i} v_{j)} \ ,     
\label{eq:h_dot} \\
\partial_t \alpha &=& v^i \widetilde{\nabla}_i \widetilde{N} -
		\widetilde{N}^2 \kappa \ ,              
\label{eq:alpha_dot} \\
\widetilde N^{-1}\partial_t^2\kappa_{ij} -
	\widetilde N\widetilde\nabla^k\widetilde\nabla_k\kappa_{ij}
        = &-& 4 \widetilde\nabla_{(i}\kappa^k_{\ j)} \widetilde\nabla_k 
	\widetilde N + \widetilde N^{-1} \kappa_{ij} \widetilde\nabla^k 
        \widetilde N \widetilde\nabla_k \widetilde N
        + 3 \widetilde\nabla^k \widetilde N \widetilde\nabla_k 
        \kappa_{ij}
        \nonumber\\
        &+& \kappa_{ij} \widetilde\nabla^k \widetilde\nabla_k \widetilde{N}
        - 2 \kappa^k_{\;(i} \widetilde\nabla_{j)} \widetilde\nabla_k 
	\widetilde{N}
        - 2 \widetilde{N}^{-1} \kappa^k_{\;(i} \widetilde\nabla_{j)} 
        \widetilde{N} \widetilde\nabla_k \widetilde{N}
        + 2 \kappa \widetilde\nabla_i \widetilde\nabla_j \widetilde{N}  
        \nonumber\\
        &+& 4 \partial_{(i} \kappa \partial_{j)} \widetilde{N}
        + 2 \widetilde{N}^{-1} \kappa \widetilde\nabla_i \widetilde{N} 
        \widetilde\nabla_j \widetilde{N}
        - 2 \widetilde{N} \widetilde{R}_{k(i} \kappa^k_{\ j)}
        - 2 \widetilde{N} \widetilde{R}_{kijm} \kappa^{km} \ ,
\label{eq:kappawave}
\end{eqnarray}
\end{mathletters}
where $\kappa \equiv \kappa^i_{\ i}$ and the tilde denotes a 
spatial quantity defined in terms of the background metric, 
$\widetilde{g}_{i j}$. Note that the wave equation for $\kappa_{i j}$
involves only the background lapse and curvature.

\subsection{Angular decomposition}
\label{subsec:angdecomp}

	We can further simplify the evolution equation
(\ref{eq:kappawave}) by separating out the angular dependence, thus
reducing it to a set of 1D equations. We accomplish this by expanding the
extrinsic curvature in Regge--Wheeler tensor spherical harmonics
\cite{rw57} and substituting this expansion into
(\ref{eq:kappawave}). Using the notation of Moncrief \cite{Moncrief}
we express the expansion as
\begin{eqnarray}
\kappa_{i j} =  a_\times (t,r) (\hat e_1)_{i j} + 
              r b_\times (t,r) (\hat e_2)_{i j} + 
 	     \widetilde N^{-2} a_+ (t,r) (\hat f_2)_{i j} + 
                   r b_+ (t,r) (\hat f_1)_{i j} + 
	\hskip 1.0truecm \nonumber\\
	        r^2 c_+ (t,r) (\hat f_3)_{i j} &+&
		r^2 d_+ (t,r) (\hat f_4)_{i j} \ ,
\label{eq:kappa_expand}
\end{eqnarray}
where $(\hat e_1)_{i j},\cdots,(\hat f_4)_{i j}$ are the Regge--Wheeler
harmonics, which are functions of $(\theta,\phi)$ and have suppressed
angular indices $(\ell,m)$ for each mode. The odd-parity multipoles
($a_\times$ and $b_\times$) and the even-parity multipoles ($a_+$,
$b_+$, $c_+$, and $d_+$) also have suppressed indices for each angular
mode and there is an implicit sum over all modes in (\ref{eq:kappa_expand}). 
The six multipole amplitudes correspond to the six components of 
$\kappa_{i j}$. However, using the linearized momentum constraints

\begin{equation}
\widetilde{\nabla}_j (\kappa^j_{\ i} - \delta^j_{\ i} \kappa) = 0 \ ,
\label{eq:MomCon}
\end{equation}
we reduce the number of independent components of $\kappa_{i j}$ to
three. An important relation is also obtained through the wave
equation for $\kappa$, whose multipole expansion is simply given by
$\kappa = h(t,r) Y_{_{\ell m}}$ where $Y_{_{\ell m}}(\theta,\phi)$ is
the standard scalar spherical harmonic and again there is an implicit
sum over suppressed indices $(\ell,m)$. Using this expansion, in
conjunction with the momentum constraints (\ref{eq:MomCon}), we derive
a set of radial constraint equations which relate the dependent
amplitudes $(b_\times)_{_{\ell m}}$, $(b_+)_{_{\ell m}}$,
$(c_+)_{_{\ell m}}$ and $(d_+)_{_{\ell m}}$ to the three independent
amplitudes $(a_\times)_{_{\ell m}}$, $(a_+)_{_{\ell m}}$, $(h)_{_{\ell
m}}$:

\begin{mathletters}
\label{eq:Constraints}
\begin{eqnarray}
(b_\times)_{_{\ell m}} &=& -\frac{1}{(\ell+2)(\ell-1)}
		[(1+3\widetilde N^2) + 
		2\widetilde N^2 r\;\partial_r]\; 
		(a_\times)_{_{\ell m}} \ , \\
(b_+)_{_{\ell m}} &=& \frac{1}{\ell(\ell+1)} [(3+r\partial_r)\; 
		(a_+)_{_{\ell m}} - (1+r\partial_r)\; 
		(h)_{_{\ell m}}] \ ,        \\
(c_+)_{_{\ell m}} &=& \frac{1}{2(\ell+2)(\ell-1)} 
		\{2(1-\ell-\ell^2)\; (a_+)_{_{\ell m}} - 
		2\; (h)_{_{\ell m}} +
		\ell(\ell+1) [(1+5\widetilde N^2) + 
		2\widetilde N^2 r\; \partial_r]\; (b_+)_{_{\ell m}}\} \ , \\
(d_+)_{_{\ell m}} &=& \frac{1}{\ell(\ell+1)}[(a_+)_{_{\ell m}} 
		+ 2 (c_+)_{_{\ell m}} - (h)_{_{\ell m}}] \ ,
\end{eqnarray}
\end{mathletters}
for each $(\ell,m)$ mode.

Substituting (\ref{eq:kappa_expand}) into (\ref{eq:kappawave}) and
using the constraint equations (\ref{eq:Constraints}), we
obtain a set of linearized radial wave equations for each independent
amplitude.  For each $(\ell,m)$ mode we have one odd-parity
equation

\begin{eqnarray}
\Biggl\{ \partial^2_t - \widetilde{N}^4 \partial^2_r - 
	\frac{2}{r}\widetilde{N}^2 \partial_r 
        - \frac{2 M}{r^3} \left(1 - \frac{3 M}{2 r} \right) + 
	\widetilde{N}^2 \left[ \frac{\ell(\ell+1)}{r^2} - 
	\frac{6 M}{r^3} \right] \Biggr\} 
        (a_\times)_{_{\ell m}} = 0 \ ,
\label{oddwave}
\end{eqnarray}
and two coupled even-parity equations,
\begin{eqnarray}
	\hskip 0.0truecm \Biggl[ \partial^2_t - \widetilde{N}^4 
	\partial^2_r - \frac{6}{r}\widetilde{N}^4 \partial_r
	+\widetilde{N}^2 \frac{\ell(\ell+1)}{r^2} - \frac{6}{r^2} +
	\frac{14M}{r^3}-\frac{3M^2}{r^4}
	\Biggr] (a_+)_{_{\ell m}} + \hskip 6.0truecm
	\nonumber\\ 
	\Biggl[\frac{4}{r} \widetilde{N}^2 \left(1 -\frac{3M}{r}\right) 
	\partial_r + \frac{2}{r^2} 
	\left(1 - \frac{M}{r} - \frac{3M^2}{r^2}\right) 
	\Biggr] (h)_{_{\ell m}} = 0 \ , 
\label{evenwave1} 
\end{eqnarray}	
\begin{eqnarray}	
	\Biggl[ \partial^2_t - \widetilde{N}^4 \partial^2_r - 
	\frac{2}{r}\widetilde{N}^2 \partial_r
	+ \widetilde{N}^2 \frac{\ell(\ell+1)}{r^2} 
	+ \frac{2 M}{r^3} -
	\frac{7 M^2}{r^4} \Biggr] (h)_{_{\ell m}} 
	- \frac{2 M}{r^3} \left(3 - \frac{7 M}{r}\right) 
	(a_+)_{_{\ell m}} = 0 \ .
\label{evenwave2}
\end{eqnarray}
These equations are related to the standard Regge--Wheeler and Zerilli
equations \cite{rw57,z70}, which can be derived in a more complete
analysis of gauge-invariant hyperbolic formulations
\cite{WorkingPaper}.

	The radial wave equations (\ref{oddwave})--(\ref{evenwave2})
for each $(\ell,m)$ mode of the independent multipole amplitudes
$(a_\times)_{_{\ell m}}$, $(a_+)_{_{\ell m}}$, $(h)_{_{\ell m}}$ form
the basis for our approach. In the perturbative region, they replace
the nonlinear Einstein equations and determine the evolution of $K_{i
j}$. They can be used to evolve, with minimal computational cost,
gravitational wave data to arbitrarily large distances from the highly
dynamical strong field region. The evolution equations for $h_{i j}$
(\ref{eq:h_dot}) and $\alpha$ (\ref{eq:alpha_dot}) can also be
integrated using the data for $K_{i j}$ computed in this region. Note
that because $h_{i j}$ and $\alpha$ evolve along the coordinate time
axis, these equations need only be integrated in the region in which
their values are desired, not over the whole region {\tt L} (these
quantities have characteristic speed zero).

\section{Numerical Implementation}
\label{sec:procimplement}

	This section is a general guide for the numerical
implementation of the Cauchy-perturbative matching method for
radiation extraction and outer boundary conditions described so far.

	Consider a 3D numerical relativity code which solves the
Cauchy problem of Einstein's equations in either the standard ADM form
\cite{ADMcode} or in the hyperbolic form \cite{Empirecode}. During
each timestep the procedure followed by our module for extracting
radiation and imposing outer boundary conditions can be summarized in
three successive steps: {\sl (1)} {\it extraction} of the independent
multipole amplitudes on {\bf E}, {\sl (2)} {\it evolution} of the radial
wave equations (\ref{oddwave})--(\ref{evenwave2}) on {\tt L} out to
the distant wave zone, {\sl (3)} {\it reconstruction} of $K_{ij}$ and
$\partial_t K_{ij}$ at specified gridpoints at the outer boundary of
{\tt N}. We now discuss in detail each of these steps:

\begin{description}	
\item {\sl(1)} \  {\it Extraction} 

	As mentioned in Sect. \ref{sec:analysis}, the extraction
2-sphere {\bf E} acts as the joining surface between the evolution of the
highly dynamical, strong field region (dark shaded area of
Fig. \ref{fig:schematic}) and the perturbative regions (light
shaded areas). At each timestep, $K_{i j}$ and $\partial_t K_{i j}$
are computed on {\tt N} as a solution to Einstein's equations. In
the test cases presented here, {\tt N} uses topologically Cartesian
coordinates, although there are no restrictions on the choice of the
coordinate system. The Cartesian components of these tensors are then
transformed into their equivalents in a spherical coordinate basis and
their traces are computed using the inverse background metric, i.e.
$H = \widetilde{g}^{i j} K_{i j}$, $\partial_t H = \widetilde{g}^{i j}
\partial_t K_{i j}$. From the spherical components of $K_{i j}$ and
$\partial_t K_{i j}$, the independent multipole amplitudes for each
$(\ell,m)$ mode are then derived by an integration over the 2-sphere:

\begin{mathletters}
\label{integrals}
\begin{eqnarray}
\label{int_1}
(a_\times)_{_{\ell m}} &=& 
	\frac{1}{\ell(\ell+1)} \int \; \frac{1}{\sin\theta}
	\left[ K_{r \phi} \; \partial_\theta - K_{r \theta} 
	\; \partial_\phi \right] \; 
	{Y^*_{_{\ell m}}} \; d\Omega \ ,\\
\label{int_2}
(a_+)_{_{\ell m}} &=& \int \; \widetilde{N}^2 \; K_{r r} 
		\; Y^*_{\ell m} \; d\Omega \ ,\\
\label{int_3}
(h)_{_{\ell m}} &=& \int \; H \; Y^*_{\ell m} d\Omega \ .
\end{eqnarray}
\end{mathletters}
Their time derivatives are computed similarly. Rather than performing
the integrations (\ref{int_1})--(\ref{int_3}) using spherical polar
coordinates, it is useful to cover {\bf E} with two stereographic
coordinate ``patches''. These are uniformly spaced two-dimensional
(2D) grids onto which the values of $K_{i j}$ and $\partial_t K_{i j}$
are interpolated using either a three-linear or a three-cubic
polynomial interpolation scheme. Each point on the 2-sphere, denoted
by spherical coordinate values $(\theta,\phi)$, corresponds to a point
$(q,p)$ on a stereographic grid whose coordinates can be combined into
a single complex number $\zeta$:

\begin{mathletters}
\begin{eqnarray}
\zeta_{_N} &\equiv& q_{_N} + i p_{_N} = 
	\tan\left(\frac{\theta}{2}\right) {\rm e}^{i \phi}\ , \\
\zeta_{_S} &\equiv& q_{_S} + i p_{_S} = \frac{1}{\zeta_{_N}} \ ,
\end{eqnarray}
\end{mathletters}
where $N$ and $S$ denote the northern ($0 \leq \theta \leq \pi/2$) and
southern ($\pi/2 \leq \theta \leq \pi$) hemispheres, respectively.  As
a result of this transformation, the integrals over the 2-sphere in
(\ref{integrals}) are computed over the stereographic patches, which
naturally avoid polar singularities (see \cite{patches} for a complete
discussion of the properties and advantages of the stereographic
coordinates). In our tests, the integrals over each patch are computed
using a second-order stereographic quadrature routine developed within
the Alliance \cite{patches}.

\item {\sl(2)} \  {\it Evolution} 

	Once the multipole amplitudes, $(a_\times)_{_{\ell m}}$,
$(a_+)_{_{\ell m}}$, $(h)_{_{\ell m}}$ and their time derivatives are
computed on {\bf E} in the timeslice $t=t_0$, they are imposed as inner
boundary conditions on the 1D grid.  Using a second-order integration
scheme (we have tested both Leapfrog and Lax-Wendroff \cite{nr}), our
module then evolves the radial wave equations
(\ref{oddwave})--(\ref{evenwave2}) for each $(\ell, m)$ mode forward
to the next timeslice at $t=t_1$. The outer boundary of the 1D grid is
always placed at a distance large enough that background field and
near-zone effects are unimportant, and a radial Sommerfeld condition
for the wave equations (\ref{oddwave})--(\ref{evenwave2}) can be
imposed there. Of course, the initial data on {\tt L} must be
consistent with the initial data on {\tt N}. This can either be
imposed analytically or determined by applying the aforementioned
extraction procedure to the initial data set at each gridpoint of
{\tt L} in the region of overlap with {\tt N}. In the latter
case, initial data outside the overlap region can be set by
considering the asymptotic fall-off of each variable.  It should be
noted that in the Cauchy-characteristic matching approach initial data
also must be set in the characteristic hypersurfaces and, for
realistic sources like binary black holes, will necessarily be
approximate.

\item {\sl(3)} \  {\it Reconstruction and Matching}

	From the perturbative data evolved to time $t_1$, outer boundary
values for {\tt N} can now be computed. The procedure for doing this
differs depending on whether a hyperbolic or an ADM formulation of
Einstein's equations is used by the 3D ``interior code''. For a
hyperbolic code (cf. \cite{Empirecode}), it is necessary to provide
boundary data for $K_{ij}$ and $\partial_t K_{i j}$. For an ADM code
(cf.  \cite{ADMcode}), on the other hand, outer boundary data only for
$K_{i j}$ are necessary, since the interior code can calculate $g_{i
j}$ at the outer boundary by integrating in time the boundary values
for $K_{i j}$. In either case, if outer boundary values for the lapse
$N$ are needed [e.g. for integrating harmonic slicing condition
(\ref{eq:harmonic})], these can be computed by the perturbative module
or by integration of $H$ at the boundary.

	In order to compute $K_{i j}$ at an outer boundary point of
{\tt N} (or any other point in the overlap between {\tt N} and
${\cal P}$ \cite{prd2}), it is necessary to reconstruct $K_{i j}$ from
the multipole amplitudes and tensor spherical harmonics.  The
Schwarzschild coordinate values $(r,\theta,\phi)$ of the relevant
gridpoint are first determined. Next, $(a_\times)_{_{\ell m}}$,
$(a_+)_{_{\ell m}}$, and $(h_{_{\ell m}})$ for each $(\ell,m)$ mode
are interpolated to the radial coordinate value of that point. The
dependent multipole amplitudes $(b_\times)_{_{\ell m}}$,
$(b_+)_{_{\ell m}}$, $(c_+)_{_{\ell m}}$, and $(d_+)_{_{\ell m}}$ are
then computed using the constraint equations (\ref{eq:Constraints}).
Finally, the Regge--Wheeler tensor spherical harmonics $(\hat{e}_1)_{i
j}$--$(\hat{f}_4)_{i j}$ are computed for the angular coordinates
$(\theta,\phi)$ for each $(\ell,m)$ mode and the sum in equation
(\ref{eq:kappa_expand}) is performed. This leads to the reconstructed
component of $\kappa_{i j}$ (and therefore $K_{i j}$); a completely
analogous algorithm is used to reconstruct $\partial_t K_{i j}$.

	It is important to emphasize that this procedure allows one to
compute $K_{i j}$ at any point of {\tt N} which is covered by the
perturbative region. As a result, the numerical module can reconstruct
the values of $K_{ij}$ and $\partial_t K_{i j}$ on a 2-surface of
arbitrary shape, or any collection of points outside of {\bf E}.
\end{description}

	Numerical implementation of this method is rather
straightforward.  Very few modifications to a standard 3D numerical
relativity code are necessary in order to allow for the simultaneous
evolution of the highly dynamical region and of the perturbative
one. Because of the use of numerically inexpensive integration of 1D
wave equations, implementation of this module provides gravitational
wave extraction {\it and} stable outer boundary conditions with only
minimal additional computational cost.

	Finally, it should be noted that, in practice, we may not know
{\em a priori} if the Schwarzschild-perturbative approximation is
valid near the outer boundary of a given numerical relativity
simulation.  Through experimentation, however, it is possible to test
the validity of the approximation. This can be done, for instance, by
extracting data at different radii and comparing the waveforms
computed at the outer sphere with those evolved from the inner
sphere. This makes it possible to determine if the neglected terms in
the approximation have a significant effect. At any point in the
overlap region between {\tt N} and ${\cal P}$, it is possible to
reconstruct gravitational wave data and compare these values with
those computed by the full nonlinear evolution.

\section{Numerical tests}
\label{sec:analytic_tests}

	In order to establish the accuracy and convergence properties
of our code we have studied the propagation of linear waves on
a Minkowski background ($M=0$) . This is a natural first test since we can
compare each stage of the numerical procedure described in Section
\ref{sec:procimplement} against a known analytic solution
\cite{Burke,Teukolsky}.

	In these tests we assign analytic values to each gridpoint of
{\tt N} at every time step. This allows us to study the accuracy
and convergence properties of the module independently of any errors
which may develop in a 3D numerical evolution of linear waves.
Elsewhere \cite{prd2}, we will present results of tests of this module
running with a full 3D evolution code (i.e. the interior code of the
Alliance \cite{ADMcode}), with emphasis on the issues of stability of
the outer boundary and accuracy of extracted waveforms.

	We have considered analytic data for $\ell=2$, $m=0$
even-parity linear waves, initially modulated by a Gaussian envelope
with amplitude $A=10^{-6}$ and width parameter $b=1$. These waves are
time-symmetric at $t=0$ and thus have ingoing and outgoing parts.  The
3D grid is vertex-centered with extents $(x,y,z) \in [-4,4]$ and
resolutions ranging from $(17)^3$ to $(129)^3$ points [corresponding
to $(16)^3$ and $(128)^3$ zones, respectively]. The resolution of the
stereographic coordinate patches corresponds to the resolution of
{\tt N} and therefore ranges from $(16)^2$ to $(128)^2$ zones on
each hemisphere. For the specific tests presented here, {\bf E} is located
at a radius $r_{_E}=3$ and similar results have been obtained also for
$r_{_E}=0.5,1.0,1.5,2,2.5,3.5$. In fact, on a flat background
spacetime and for weak waves on Schwarzschild-like backgrounds, the
perturbative approximation is valid throughout the 3D domain and the
position of {\bf E} is thus arbitrary.

\begin{figure}
\epsfxsize=10cm
\begin{center}
\leavevmode
\epsffile{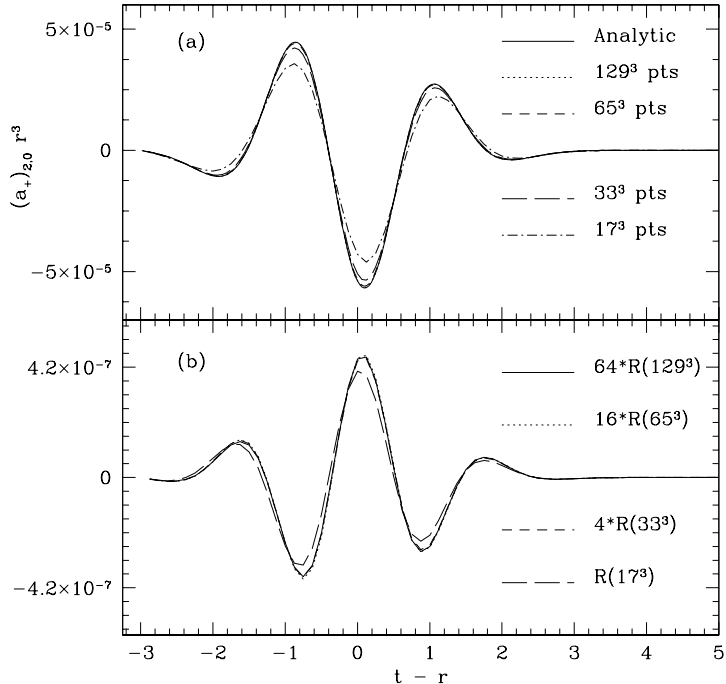}
\vskip 1.0truecm 
\end{center}
\caption[fig2]{\label{fig:extraction} 
	(a) Timeseries of the multipole amplitude $(a_+)_{_{2 0}}$
	extracted at a 2-sphere of radius $r_{_E}=3$ for various grid
	resolutions. The amplitude is scaled by $r^3$ to compensate
	for the radial fall-off. (b) Residuals of the leading order
	error term for different grid resolutions differing by a
	factor of 2. The residuals are multiplied by 4, 16, and 64 in
	order to make the errors comparable. If no higher-order terms
	were present, all of the curves would coincide.}
\end{figure}

	Since these waves are traceless and even-parity, with pure
$\ell=2$, $m=0$ angular dependence, the only non-zero independent
multipole amplitude we expect to find at {\bf E} is $(a_+)_{_{2
0}}$. Diagram (a) of Fig. \ref{fig:extraction} shows plots of
$(a_+)_{_{2 0}}$ extracted at $r_{_E}=3$ as a function of $t-r$ for
various resolutions of {\tt N}. The amplitude is scaled by $r^3$ to
compensate for the radial fall-off.

	The curves in diagram (a) clearly show that the extracted
waveform approaches the analytic value (denoted by a solid line) as
the resolution is increased. However, in order to establish the exact
rate at which the computed solution approaches the analytic one, we
have also performed convergence tests. These tests are designed 
to check that no coding error has been made and that the numerical
scheme employed in the solution is providing results at the expected
accuracy. While there are a number of different ways to perform these
tests, we have exploited the knowledge of an analytic solution and
computed the residuals $R$ between the computed solution
${\mathcal F}_c$ and the analytic one ${\mathcal F}_a$ as a function
of the resolution (or, equivalently, of the number of gridpoints). For
a second-order accurate numerical scheme (as the one used here) on a
uniform cubical grid, we expect the residuals to follow the simple law

\begin{equation}
R(N^3) = {\mathcal F}_c - {\mathcal F}_a = 
O(h^2) \ ,
\end{equation}
where $O(h^2)$ contains the second and higher order error terms and
$h=L/(N-1)$ is the grid resolution, with $L$ being the spatial
dimension of the grid. If the numerical computation is second-order
accurate and a number of simulations with different grid resolutions,
each differing by a factor 2, are performed, we should expect the
residual to fall quadratically to zero. Diagram (b) of
Fig. \ref{fig:extraction} shows this is indeed the case; there, we
have multiplied the residuals obtained with different resolutions by
the coefficients that make the leading order error terms comparable. The
good overlapping of the different curves is an indication that a
second-order convergence has been achieved.

\begin{figure}
\epsfxsize=10cm
\begin{center}
\leavevmode
\epsffile{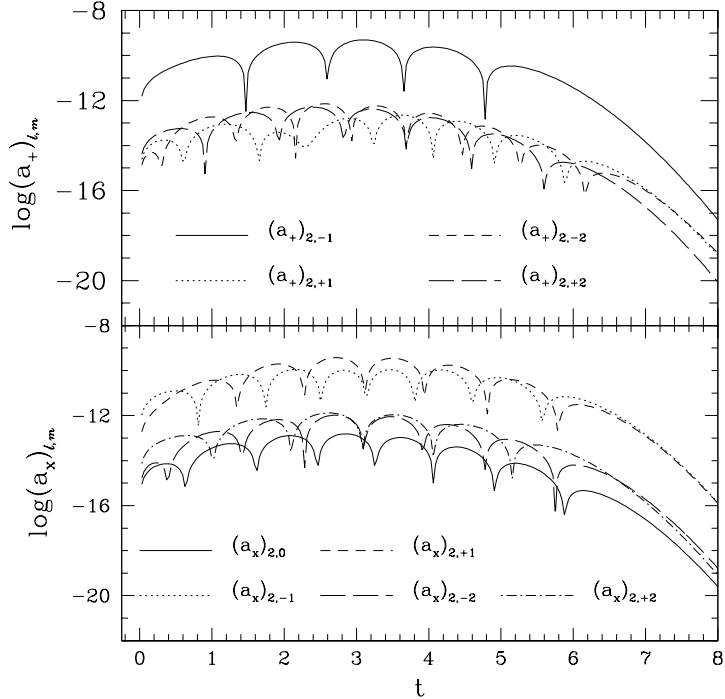}
\vskip 1.0truecm 
\end{center}
\caption[fig3]{\label{fig:zeromodes}
	Timeseries of the analytically vanishing even
	$(a_+)_{_{\ell m}}$ and odd $(a_{\times})_{_{\ell m}}$
	parity multipole amplitudes. The extraction is made at
	a 2-sphere of radius $r_{_E} = 3$ and {\tt N} has 
	$(65)^3$ points.}
\end{figure}

	The accuracy of this extraction procedure can also be tested
by examining the waveforms for the other multipole amplitudes computed
which analytically vanish. Figure \ref{fig:zeromodes} shows plots of
several even-parity (upper diagram) and odd-parity (lower diagram)
amplitudes computed at the extraction 2-sphere for an resolution in
{\tt N} of $(65)^3$ points. As a result of numerical truncation
error introduced in the extraction procedure, these modes are not
exactly zero. However, even the largest amplitude mode is over three
orders of magnitude smaller than the only analytically non-vanishing
independent amplitude $(a_+)_{_{2 0}}$.  Moreover, all of these
amplitudes are second-order convergent to zero as the resolution is
increased.  Similar considerations apply also for the $(h)_{_{\ell
m}}$ multipole amplitudes: although the data is analytically
traceless, very small $(h)_{_{\ell m}}$ modes are extracted at the
2-sphere. These modes, which we will not show here, are the order of
round-off error (approximately $10^{-22}$ for these tests) and may be
considered as effectively zero.

\begin{figure}
\epsfxsize=10cm
\begin{center}
\leavevmode
\epsffile{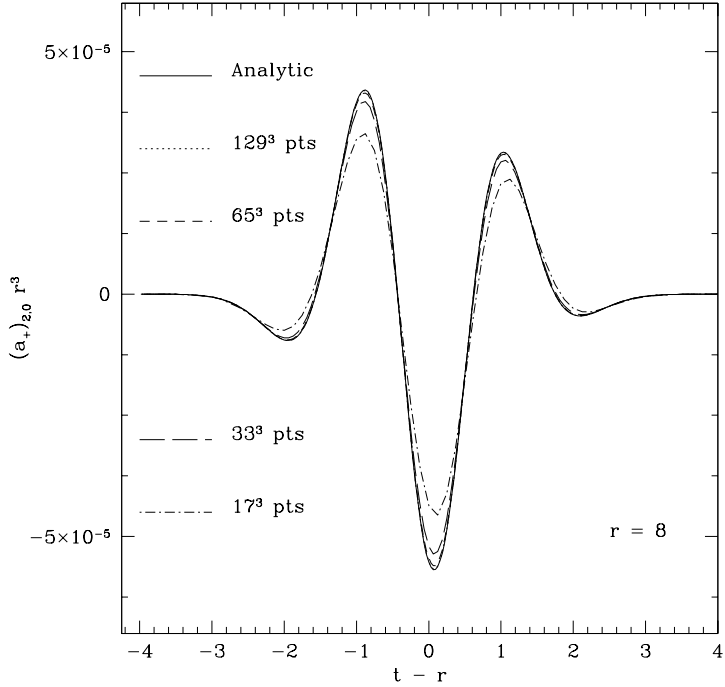}
\vskip 1.0truecm 
\end{center}
\caption[fig4]{\label{fig:evolved}
	Timeseries of $(a_+)_{_{2 0}}$ evolved to $r=8$ for various 
	grid resolutions. Here also, the amplitude is scaled by $r^3$ to
	compensate for the radial fall-off.}
\end{figure}

	Next, we consider the accuracy of the evolution in the
perturbative region of the extracted amplitudes. The time integration
of (\ref{oddwave})--(\ref{evenwave2}) on {\tt L} is performed using
a Leapfrog integration scheme with a spatial resolution adjusted so
that the timesteps in {\tt N} and {\tt L} are identical. This
imposes a relation involving the gridspacing of {\tt N} and the
ratio of Courant factors for {\tt N} and {\tt L}. Such a choice
ensures a correspondence between resolutions of {\tt N} and ${\cal
P}$ Fig. \ref{fig:evolved} shows plots of $(a_+)_{_{2 0}}$ evolved to
a radius $r=8$ from the extracted signal at $r=r_{_E}=3$. Different
curves correspond to different resolutions and show the convergence to
the analytic solution. The outer boundary of {\tt L} is located at
$r=33$, where outgoing wave Sommerfeld conditions are imposed. For
radial scalar wave equations, this represents a very good
approximation which has been shown to be both accurate and stable.

\begin{figure}
\epsfxsize=10cm
\begin{center}
\leavevmode
\epsffile{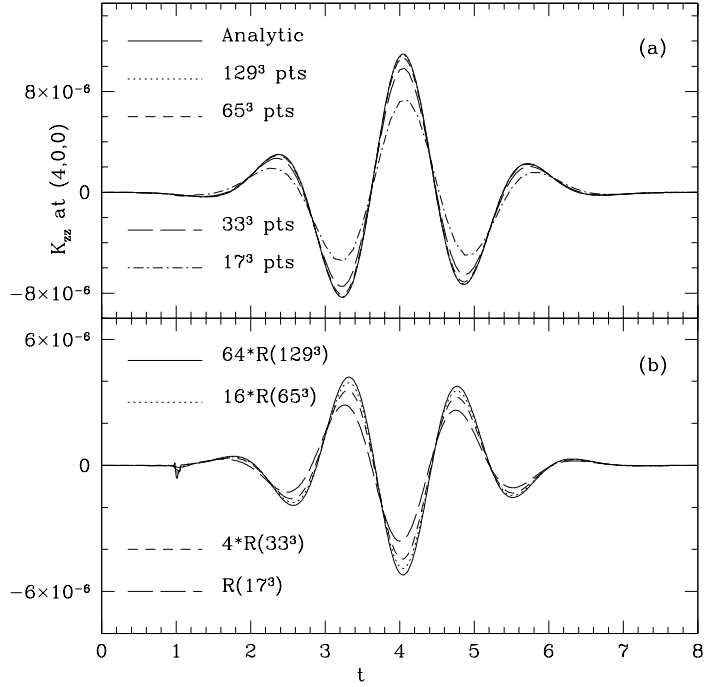}
\vskip 1.0truecm 
\end{center}
\caption[fig5]{\label{fig:kzzpoint} 
	(a) Timeseries of the reconstructed values for $K_{z z}$ at
	the grid point $(4,0,0)$ for various grid resolutions.  (b)
	Residuals of the leading order error term for different grid
	resolutions differing by a factor of 2
	(cf. Fig. \ref{fig:extraction}).}
\end{figure}

	Finally, we consider the accuracy in the reconstruction of the
outer boundary data. Since we are using analytic data in {\tt N},
we can only compare the outer boundary data with the analytic ones. In
a forthcoming paper \cite{prd2}, where we will make use of a numerical
solution of Einstein's equations, we will also discuss the issues of
stability related to the use of a Cauchy-perturbative matching method.
For conciseness we consider here reconstructed outer boundary data
only for $K_{i j}$; the reconstruction of $\partial_t K_{i j}$ follows
analogously. Diagram (a) of Fig. \ref{fig:kzzpoint} shows the
timeseries of the reconstructed value of $K_{z z}$ computed at the
point $(x=4, y=0, z=0)$ for various resolutions and its comparison
with the analytic solution.  Also in this case, diagram (b) of
Fig. \ref{fig:kzzpoint} gives proof of the second-order convergence of
the numerical module even if, in this case, higher order error terms
become apparent with the very coarse resolution simulations [i.e. in
the case of $(17)^3$ gridpoints]. The small peak observed at $t-r
\approx 1$ is the result of a slight difference between the analytic
initial data on {\tt L} and the extracted signal at $t=0$.  This
error rapidly disappears as the resolution is increased.

	A more global measure of the accuracy and of the convergence
properties of the boundary data is obtained by computing the $L_2$
norm of the error in $K_{i j}$ as measured over the whole 3D outer
boundary. In Fig. \ref{fig:kzznorm} we plot the $L_2$ norms of $K_{zz}$ 
at successive resolutions, normalizing these differences by the factor 
which would make the plots overlap if the
convergence to analytic data were exactly second-order. Here we again
see that the desired convergence rate is achieved over the whole
boundary, particularly at finer grid resolutions.

\vbox{
\begin{figure}
\epsfxsize=10cm
\begin{center}
\leavevmode
\epsffile{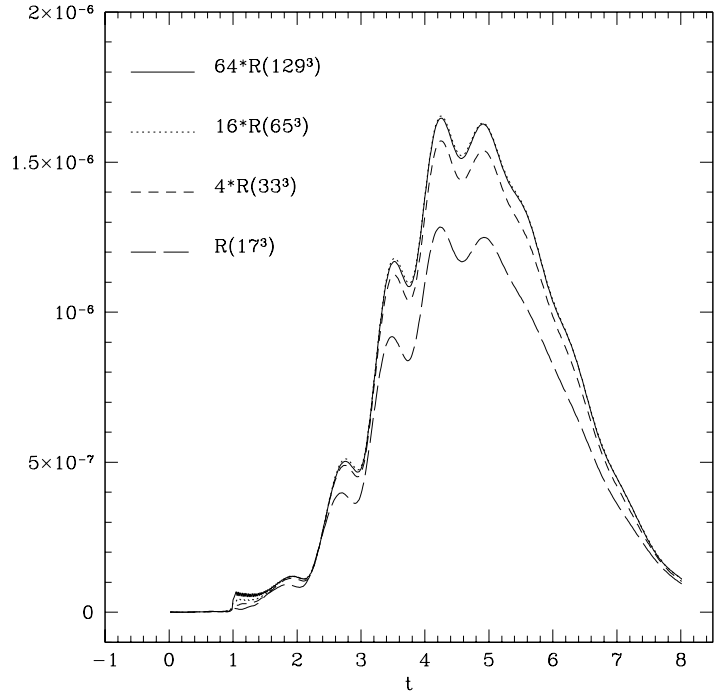}
\vskip 1.0truecm 
\end{center}
\caption[fig6]{\label{fig:kzznorm}
	Plot of $L_2$ norms of $K_{z z}$ computed over the outer 
	boundary for successive grid resolutions. The norms at 
	different grid resolutions are scaled so that they overlap if
	truly second-order convergent.}
\end{figure}
}

\section{Conclusion}
\label{sec:discussion}

	We have presented a method for matching gravitational data
computed from a 3D Cauchy evolution of Einstein's equations to a
computationally simpler evolution of radial wave equations linearized
on a Schwarzschild background. This method should be applicable to a
variety of physical problems where curvature is significant throughout
the computational domain, so long as the time-dependent fields can be
treated as linear perturbations on a spherical background. Our
approach promises to offer an accurate means of computing
asymptotically gauge-invariant waveforms at large distances from the
domain of the simulation and to provide stable, physically correct 
boundary conditions.

	We have also discussed a numerical code we have developed that
implements this procedure and can be used with a general numerical 3D
simulation, solving Einstein's equations in either the hyperbolic or
ADM formulation. This code correctly extracts waveforms from analytic
linear wave data and recomputes that data at the outer boundary of the
3D grid. A more extensive discussion of the stability properties of
this approach will be discussed in a forthcoming paper \cite{prd2}, as
well as practical issues arising from application to a real evolution
code environment.

	Our Cauchy-perturbative matching method can be extended to
more general circumstances, e.g. perturbations on axially symmetric
backgrounds or other slicings of Schwarzschild black holes.  Similar
analyses using other hyperbolic formulations\cite{CBYBanach,OtherHyp}
may also provide important insight to the physical understanding of
radiation extraction and lead to modules which work with simulations
based on these formulations.

\acknowledgments

	We wish to thank J.~Lenaghan for early work on the
construction of the perturbative outer boundary code, and A.~Anderson,
S.~L.~Shapiro and J.~W.~York for helpful discussions and careful
reading of this manuscript. We would also like to thank R.~G\'omez,
L.~Lehner, P.~Papadopoulos and J.~Winicour for providing the numerical
routines that performed the integration over the stereographic
patches. This work was supported by the NSF Binary Black Hole Grand
Challenge Grant Nos. NSF PHY 93--18152, NSF PHY 93--10083, ASC
93--18152 (ARPA supplemented). M.~E.~R. also acknowledges support from
NSF Grant No. PHY 94--13207 to the University of North Carolina.
Computations were performed at NPAC (Syracuse University) and at NCSA
(University of Illinois at Urbana-Champaign).

\end{document}